\documentclass[preprint,12pt]{elsarticle}

\usepackage{amssymb}
\usepackage{booktabs}
\usepackage{amsmath,amsfonts}
\usepackage{times}
\usepackage{lineno}

\journal{arXiv}

\begin{document}

\begin{frontmatter}

\title{A Novel Low-Power Cache Architecture Based on 6-Transistor SRAM Cells}

\author[add1]{Naser Khatti Dizabadi \corref{cor1}}
\ead{nak5300@utulsa.edu}
\cortext[cor1]{Corresponding author}


\author[add1]{Ceyda Elcin Kaya}
\ead{cek7838@utulsa.edu}

\affiliation[add1]{organization={Department of Electrical and Computer Engineering, The University of Tulsa},
            addressline={800 S. Tucker Drive},
            city={Tulsa},
            postcode={74104},
            state={OK},
            country={USA}}


\begin{abstract}
This paper presents a low-power cache architecture based on the series interconnection of conventional 6-transistor static random-access memory (6T SRAM) cells. The proposed approach aims to reduce leakage power in SRAM-based cache memories without increasing the transistor count of the memory cell itself. In the proposed architecture, adjacent cells within a column are reconfigured in a serial topology, thereby exploiting the stacking effect to suppress leakage current, particularly during hold operation. This architectural modification requires corresponding changes to the addressing and sensing structure of the cache, including adjustments to the column organization and readout path. To evaluate the proposed method, transient simulations were carried out using Keysight ADS. The simulation results show that the proposed architecture reduces leakage power compared with the conventional SRAM interconnection scheme while preserving the use of standard 6T SRAM cells.
\end{abstract}


\end{frontmatter}


\section{Introduction}
\label{intr}

The rapid expansion of Internet of Things (IoT) systems, mobile electronics, and wearable medical devices has intensified the demand for energy-efficient hardware design. Many of these systems operate under stringent power constraints due to limited battery capacity or reliance on energy-harvesting techniques. Under such conditions, reducing power consumption is a primary design objective. Among the major components of digital systems, memory subsystems play a particularly important role in overall energy usage. Although SRAM-based cache memories are widely used to support high-speed operation, their leakage power remains a significant challenge in low-power applications.

Modern computing platforms, including computers, microcontrollers, wearable devices, and IoT nodes, rely heavily on memory components. In memory design, two performance criteria are especially important: speed and power efficiency. Within the memory hierarchy, cache memories provide fast data access, second only to registers, but they also contribute substantially to total power consumption. Because cache memories are commonly built from arrays of SRAM cells, improving SRAM energy efficiency has become an important area of research.

A variety of approaches have been proposed to reduce SRAM power consumption. The feedback-cutting 7T SRAM cell in \cite{feedback-cutting} reduces power consumption through cell-level circuit modification. A fully differential low-power 8T SRAM structure is introduced in \cite{fully-diff}, while \cite{disturb-free} proposes a single-ended disturb-free 9T SRAM cell for subthreshold operation. In \cite{IGFinFET}, a low-leakage 6T SRAM cell is developed using insulated-gate FinFET technology. A hybrid biasing and power-gating method is reported in \cite{hybrid-biasing} to reduce leakage during hold mode. In addition, \cite{data-dependent-A} and \cite{data-dependent-B} explore data-dependent SRAM design strategies for improving power efficiency. Another widely studied technique is transistor stacking, which reduces leakage current by exploiting the behavior of serially connected transistors \cite{Stacking-A, Stacking-B, Stacking-C, Stacking-D, Stacking-E}.

Despite these advances, most existing approaches focus on modifying the internal SRAM cell structure. In many cases, this results in cells containing more than six transistors, which increases area overhead and, consequently, fabrication cost. In contrast, the present work preserves the conventional 6T SRAM cell and instead focuses on architectural reconfiguration at the cache level. More specifically, adjacent SRAM cells are serially interconnected and their supply arrangement is modified so that leakage current can be reduced without significantly increasing chip area. The proposed method therefore seeks to achieve lower leakage power while retaining the compact structure of the standard 6T SRAM cell.

The remainder of this paper is organized as follows. Section 2 reviews the conventional cache architecture and the operating modes of a standard SRAM cell. Section 3 describes the proposed cache architecture and circuit topology. Section 4 presents the simulation setup and results. Finally, Section 5 concludes the paper.

\section{Conventional Cache Architecture}
\label{Conv}

\begin{figure}[] 
    \centering
    \includegraphics[width=8cm]{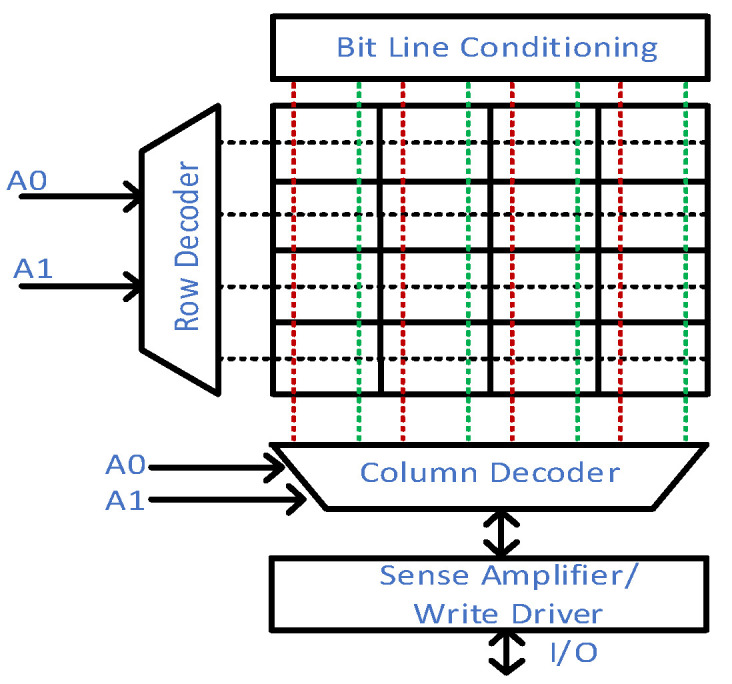}
    \caption{Conventional cache architecture \cite{General_SRAM_Architecture}}
    \label{fig-1}
\end{figure}

Figure \ref{fig-1} illustrates a conventional cache architecture consisting of row and column decoders, bit lines, and sense amplifiers. In this structure, each SRAM cell is directly connected to the global supply rails, namely VDD and GND. This straightforward arrangement enables standard read and write operations but does not take advantage of architectural leakage-reduction techniques at the array level.

\begin{figure}[] 
    \centering
    \includegraphics[width=8cm]{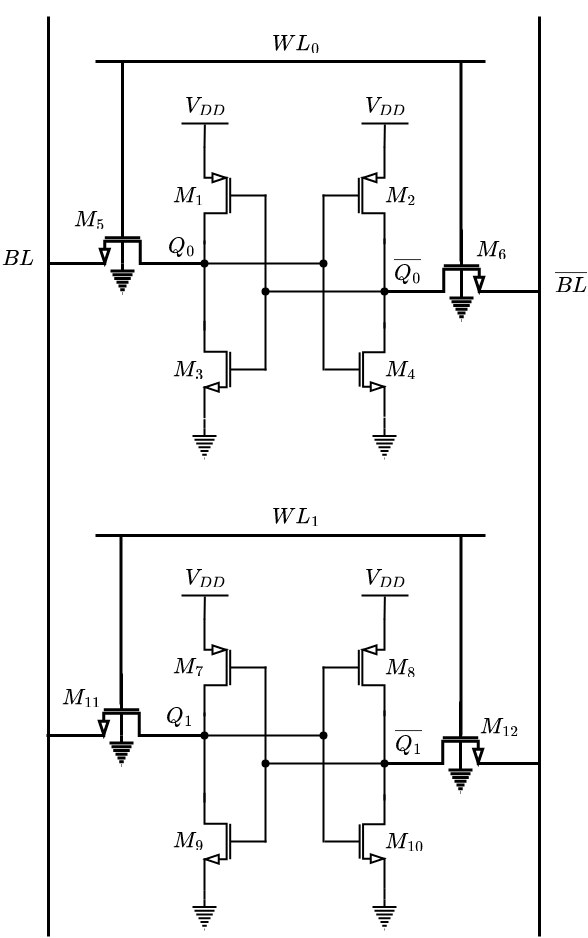}
    \caption{Conventional 6T SRAM cell}
    \label{fig-2}
\end{figure}

Figure \ref{fig-2} shows the transistor-level topology of a conventional 6T SRAM cell. The first and second cells contain core transistors M1--M4 and M7--M10, respectively. Access transistors M5, M6, M11, and M12 provide connectivity between the internal storage nodes and the bit lines, \(BL\) and \(\overline{BL}\). The word-line signals \(WL_0\) and \(WL_1\) control access to the first and second cells, respectively. As commonly described in the literature, SRAM cells operate in three primary modes: read, write, and hold \cite{Op-modes}.

\subsection{Read Mode}
During a read operation, the bit lines \(BL\) and \(\overline{BL}\) are first precharged to VDD. The word line is then asserted, turning on the access transistors and connecting the internal storage nodes to the bit lines. Depending on the stored logic value, one bit line experiences a slight discharge relative to the other. This voltage difference is sensed by the sense amplifier and interpreted as the stored data.

\subsection{Write Mode}
During a write operation, the input data and its complement are applied to \(BL\) and \(\overline{BL}\), respectively. The word line is then asserted so that the access transistors conduct and allow the bit-line values to overwrite the previous state of the cross-coupled inverters. As a result, a new logic value is stored in the cell.

\subsection{Hold Mode}
In hold mode, the word line remains deasserted, which keeps the access transistors off and isolates the cell from the bit lines. Under this condition, the cross-coupled inverters maintain the previously stored value as long as the supply voltage is preserved.

\section{Proposed Architecture and Circuit Topology}
\label{Proposed}

\begin{figure}[] 
    \centering
    \includegraphics[width=8cm]{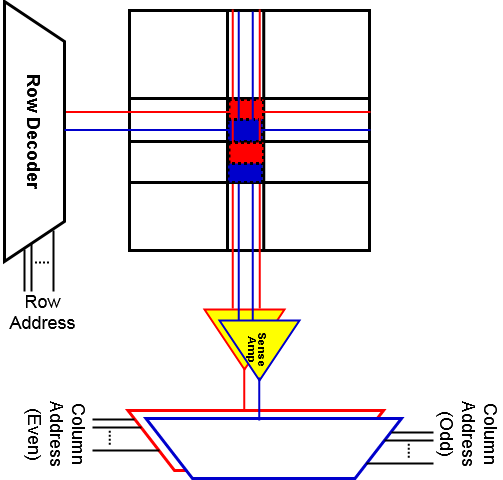}
    \caption{Proposed cache architecture}
    \label{fig-3}
\end{figure}

The proposed cache architecture is shown in Figure \ref{fig-3}. In this structure, each pair of adjacent SRAM cells in the same column is connected in series rather than being independently tied to the global supply rails. This arrangement is intended to exploit the stacking effect and thereby reduce leakage current. The row decoder remains unchanged; however, the column organization must be modified because two separate bit-line pairs are used: \(BL0\), \(\overline{BL0}\), \(BL1\), and \(\overline{BL1}\). As a result, two sense amplifiers are required per column, and the column decoder is partitioned so that the appropriate readout path can be selected.

Because two consecutive cells are connected in series, the supply arrangement must be adjusted so that both cells can operate correctly. In the proposed design, the effective voltage levels of the upper and lower cells are no longer identical to those of the conventional architecture. In particular, the output voltage range of the upper cell differs from that of the lower cell, which motivates the use of separate bit-line pairs and corresponding sensing paths. Additional precharge capacitors are also incorporated to ensure correct precharge operation under the modified voltage conditions. In this configuration, one sensing path must be capable of detecting voltage differences in an elevated operating range while still producing a standard logic-level output.

\begin{figure}[] 
    \centering
    \includegraphics[width=8cm]{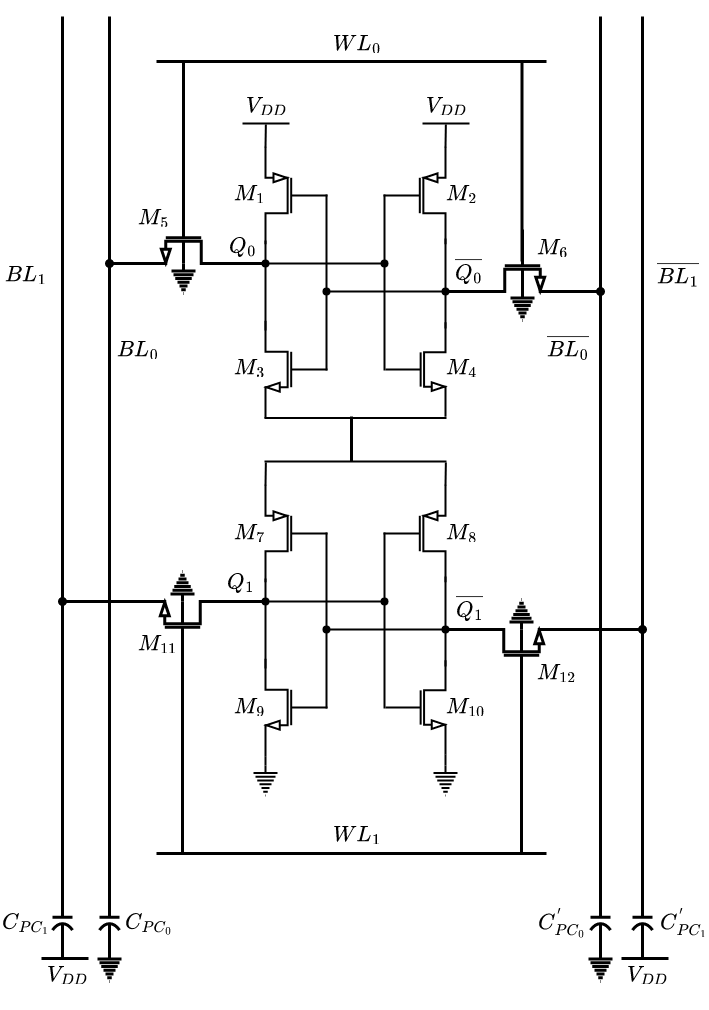}
    \caption{Main idea of the proposed SRAM cell}
    \label{fig-4}
\end{figure}

Figure \ref{fig-4} illustrates the basic principle behind the proposed cell interconnection. The lower supply node of the upper SRAM cell is connected to the upper supply node of the lower SRAM cell, thereby forming a serially connected pair. To support correct operation, the overall supply arrangement is increased accordingly. This configuration enables leakage reduction through transistor stacking, especially when the cells are in hold mode and several transistors remain off. Consequently, the leakage current of the proposed structure is expected to be lower than that of the conventional architecture.

To realize the proposed concept in practice, several additional control and support circuits are required, as shown in Figure \ref{fig-5}. The bit lines are connected to PMOS devices that provide the precharge paths. For the voltage-shifted branch, the bit lines are precharged using higher voltage levels, and the associated PMOS devices are connected accordingly. In addition, two data-enable signals, \(DEN0\) and \(DEN1\), are used to determine which bit-line pair is driven by the data input during write operation.

\subsection{Read Operation in the Proposed Topology}
The read operation begins with precharging the appropriate bit-line pair. If the upper cell is selected, \(C_{PC0}\) and \(C^{'}_{PC0}\) are charged to the required levels. After precharging is completed, the data path remains disabled by setting \(DEN0\) low, and \(WL0\) is asserted. The internal storage nodes of the upper cell then induce a differential voltage on \(BL0\) and \(\overline{BL0}\), which is detected by the corresponding sense amplifier.

If the lower cell is selected, the same sequence is applied to the second branch. In that case, \(BL1\) and \(\overline{BL1}\) are precharged, \(DEN1\) remains low, and \(WL1\) is asserted. The resulting bit-line differential is then sensed by the second amplifier. After sensing, the proper output is selected through the column decoding circuitry depending on which cell has been accessed.

\subsection{Write Operation in the Proposed Topology}
For a write operation, the desired data must first be applied to the appropriate bit-line pair. If the upper cell is selected, \(DEN0\) is asserted so that the data input drives \(BL0\) and \(\overline{BL0}\). Then \(WL0\) is enabled, allowing the access transistors to transfer the bit-line values into the upper SRAM core.

Similarly, if the lower cell is selected, \(DEN1\) is asserted to drive \(BL1\) and \(\overline{BL1}\), after which \(WL1\) is enabled. This allows the lower cell to capture the new logic value from the bit lines.

\subsection{Hold Operation in the Proposed Topology}
In hold mode, both cells must remain electrically isolated from their corresponding bit lines so that the stored data is preserved. To achieve this, \(WL0\) and \(WL1\) are both connected to GND, turning off the access transistors in both branches. The data path is also disabled by setting \(DEN0\) and \(DEN1\) low. Finally, the precharge operation is disabled so that no unnecessary charging activity occurs during standby. Under these conditions, the serial configuration benefits from the stacking effect, which helps reduce leakage current relative to the conventional topology.


\section{Simulation Results}
\label{Sim}

The proposed topology was simulated using the Keysight Advanced Design System (ADS). A pulse generator was used to provide representative input data, and suitable control voltages were applied to the word-line, data-enable, and precharge signals. Transient simulations were then carried out to examine the electrical behavior of the proposed structure.

\begin{figure}[] 
    \centering
    \includegraphics[width=8cm]{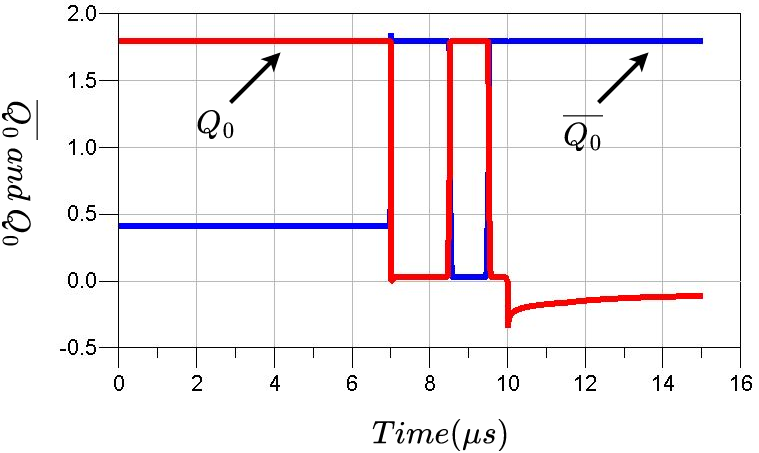}
    \caption{Output voltages of the lower core in the proposed SRAM cell}
    \label{fig-6}
\end{figure}

\begin{figure}[] 
    \centering
    \includegraphics[width=8cm]{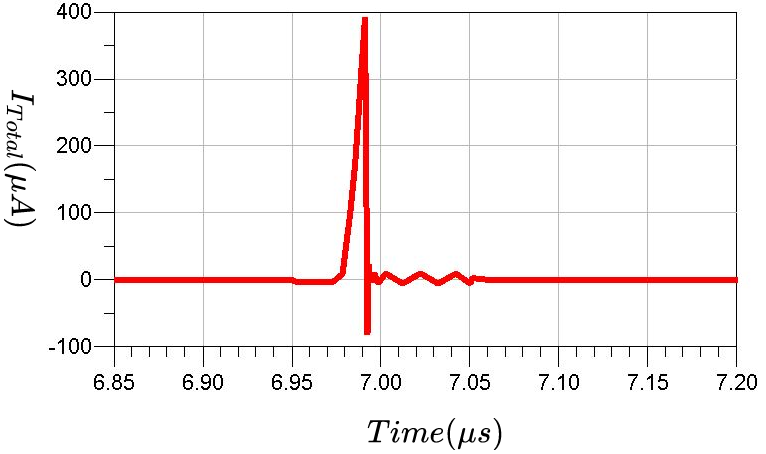}
    \caption{Total current drawn by the proposed SRAM cell}
    \label{fig-7}
\end{figure}

\begin{figure}[] 
    \centering
    \includegraphics[width=8cm]{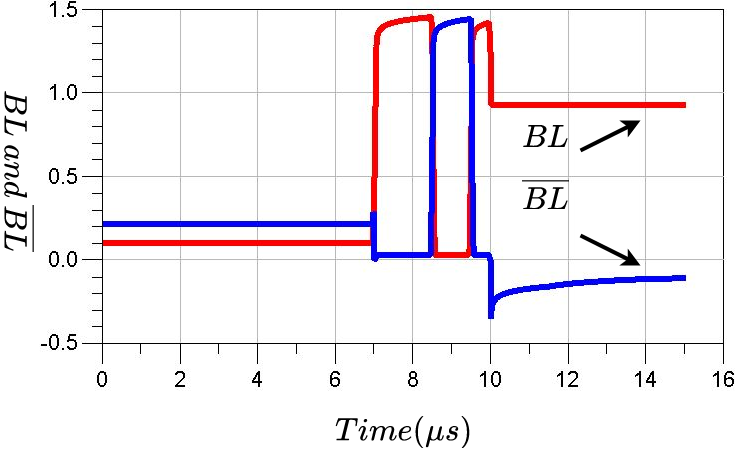}
    \caption{Simulated bit lines of the proposed SRAM cell}
    \label{fig-8}
\end{figure}

\begin{figure}[] 
    \centering
    \includegraphics[width=8cm]{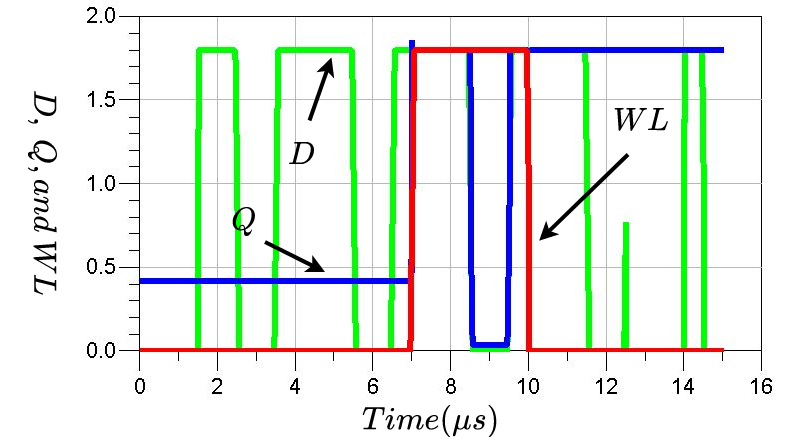}
    \caption{Simulated data line, cell output, and word-line signal of the proposed SRAM topology}
    \label{fig-9}
\end{figure}

Figure \ref{fig-6} shows the simulated output voltages of the lower SRAM core. Figure \ref{fig-7} presents the total current drawn by the proposed circuit. The bit-line behavior for the lower cell is shown in Figure \ref{fig-8}. Finally, Figure \ref{fig-9} illustrates the relationship among the input data signal, the cell output, and the word-line control during operation. These results confirm that the proposed topology functions correctly under the simulated conditions and support the intended low-leakage behavior of the design.

\section{Conclusion}
\label{Con}

This paper presented a cache architecture for leakage-power reduction based on the series interconnection of conventional 6T SRAM cells. Unlike many prior low-power SRAM approaches that modify the internal cell structure and increase transistor count, the proposed method preserves the standard 6T cell while introducing an architectural reconfiguration at the cache level. By exploiting the stacking effect in serially connected cells, the proposed topology reduces leakage power, particularly in hold mode. Simulations performed in Keysight ADS and demonstrated the feasibility of the proposed approach and its potential for low-power cache design. The results indicate that the proposed architecture can provide leakage-power savings relative to the conventional SRAM interconnection scheme while maintaining a compact cell structure.

\bibliographystyle{elsarticle-num}
\bibliography{MyBibDatabase}

\end{document}